\documentclass[a4paper]{jpconf}
\usepackage{graphicx}

\begin{document}
\title{SUSY dark matter annihilation in the Galactic halo}
\author{Veniamin Berezinsky$^{1,2,3}$, Vyacheslav Dokuchaev$^{1}$, Yury Erohenko$^{1}$}
\address{$^{1}$ Institute for Nuclear Research of the Russian Academy of Sciences, \\
prospekt 60-letiya Oktyabrya 7a, 117312 Moscow, Russia}
\address{$^{2}$ Gran Sasso Science Institute (INFN), viale F. Crispi 7, 67100 L'Aquila, Italy}
\address{$^{3}$ INFN/Laboratori Nazionali Gran Sasso, ss 17bis km 18+910, 67100 Assergi, Italy}
\ead{berezinsky@lngs.infn.it, dokuchaev@inr.ac.ru, eroshenko@inr.ac.ru}

\begin{abstract}
Neutralino annihilation in the Galactic halo is the most definite observational signature proposed for indirect registration of the SUSY Dark Matter (DM) candidate particles. The corresponding annihilation signal (in the form of gamma-rays, positrons and antiprotons) may be boosted for one or three orders of magnitude due to the clustering of cold DM particles into the small-scale and very dense self-gravitating clumps. We discuss the formation of these clumps from the initial density perturbations and their successive fate in the Galactic halo. Only a small fraction of these clumps, $\sim0.1$~\%, in each logarithmic mass interval $\Delta\log M\sim1$ survives the stage of hierarchical clustering. We  calculate the probability of surviving the remnants of dark matter clumps in the Galaxy by modelling the tidal destruction of the small-scale clumps by the Galactic disk and stars. It is demonstrated that a substantial fraction of clump remnants may survive through the tidal destruction during the lifetime of the Galaxy. The resulting mass spectrum of survived clumps is extended down to the mass of the core of the cosmologically produced clumps with a minimal mass. The survived dense remnants of tidally destructed clumps provide an amplification (boosting) of the annihilation signal with respect to the diffuse DM in the Galactic halo. We describe the anisotropy of clump distribution caused by the tidal destruction of clumps in the Galactic disk.
\end{abstract}

\section{Introduction}
\pagenumbering{arabic}

The stable SUSY particle like neutralino is a promising candidate for the enigmatic Dark Matter (DM) particle. A primordial power-law spectrum of density fluctuations in the DM ranges from the largest scales above the scales of superclusters of galaxies to the smallest sub-stellar scales according to prediction of inflation models. This permits to predict the properties of smallest DM structures from the known Cosmic Microwave Background (CMB) fluctuations at large scales. The most bright indirect signature of the SUSY DM particles is their annihilation, and this process could be boosted inside the dense DM clumps in the galactic halo.

The cosmological formation and evolution of small-scale DM clumps have been studied in numerous works \cite{SchSchWid99,SchHof00,bde03,ZTSH,GreHofSch05,
bde06,DieKuhMad06,Green07,bertsh,AngZha07,bde07,GioPieTor07,
bdeks10,bdeks10b,bde11,bde13}. The minimum mass of clumps (the cutoff of the mass spectrum), $M_{\rm min}$ is determined by the collision and collisionless damping processes (see, e.\,g., \cite{GreHofSch05} and references therein). Additionally the cutoff of mass spectrum is influenced by the acoustic absorption \cite{weinberg} at the time of kinetic decoupling of the SUSY DM particles \cite{bino} and also by the horizon-scale perturbation modes \cite{LoeZal05}. The low-mass cut-off of the clump mass-spectrum accompanies the process of decoupling. It starts when DM particles  coupled strongly with surrounding plasma in the growing density fluctuations. The smearing of the small-scale fluctuations is due to the collision damping occurring just before decoupling, in analogy with the Silk damping \cite{Silk68}.  It occurs due to diffusion of DM particles from a growing fluctuation, and only the small-scale fluctuations can be destroyed by this process. The corresponding diffusive cut-off $M_{\rm min}^{\rm diff}$ is very small. As coupling
becomes weaker, the larger fluctuations are destroyed and $M_{\rm min}$ increases. One may expect that the largest value of $M_{\rm min}$ is related to a free-streaming regime. However, as recent calculations show \cite{bertsh}, the largest $M_{\rm min}$ is related to some friction between DM particles and cosmic plasma similar to the Silk damping. The predicted minimal clump masses range from very low values, $M_{\rm min}\sim10^{-12}{\rm M}_\odot$ \cite{zvg}, produced by diffusive escape of DM particles, up to $M_{\rm min}\sim10^{-4}{\rm M}_\odot$, caused by acoustics oscillations \cite{LoeZal05} and quasi-free-streaming with limited friction \cite{bertsh}. In the case of the Harrison-Zeldovich spectrum of primordial fluctuations with CMB normalization the first Earth-mass small-scale DM clumps are formed at redshift $z\sim60$ (for $2\sigma$ fluctuations) with a mean density $7\times10^{-22}$~g~cm$^{-3}$, virial radius $6\times10^{-3}$~pc and internal velocity dispersion $80$~cm~s$^{-1}$ respectively. Only very small fraction of these clumps survives the early stage of tidal destruction during the hierarchial clustering \cite{bde03}. Nevertheless these survived clumps may provide the major contribution to the annihilation signal in the Galaxy
\cite{bde03,bde06,KamKou08,PieBerBra07,AndKom06}. At a high redshift the SUSY neutralinos, considered as DM particles, may cause the efficient heating of the diffuse gas \cite{MyeNus07} due to annihilation in the dense clumps.

One of the unresolved problem of DM clumps is a value of the central density or core radius. Numerical simulations give a nearly power density profile of DM clumps. Both the Navarro-Frenk-White (NFW) and Moore profiles give formally a divergent density in the clump center. A theoretical modeling of the clump formation \cite{ufn1,ufn2,ufn3} predicts a power-law profile of the internal density of clumps
\begin{equation}
 \rho_{\rm int}(r)=
 \frac{3-\beta}{3}\,\bar\rho\left(\frac{r}{R}\right)^{-\beta},
 \label{rho}
\end{equation}
where $\bar\rho$ and $R$ are  the mean internal density and a radius of clump, respectively,  $\beta\simeq1.8-2$ and $\rho_{\rm int}(r)=0$ at $r>R$. A near isothermal power-law profile
(\ref{rho}) with $\beta\simeq2$ has been  recently obtained in numerical simulations of small-scale clump formation \cite{DieMooSta05}. In \cite{bde03} the core radius  $x_c\simeq 0.3\nu^{-2}$ has been obtained, where $\nu$ is a relative height of the fluctuation density peak in units of dispersion at the time of energy-matter equality.  This value is a result of  the influence of tidal forces on the motion of DM particles in the clump at stage of formation. This estimate may be considered as an upper limit for the core radius or as the break-scale in the density profile, e.\,g., a characteristic scale in the Navarro-Frenk-White profile. It could be that a real core radius, where the density ceases to grow, is determined by the relaxation of small-scale perturbations inside the forming clump \cite{DorLuk}. Another mechanism for core formation arises in the `meta-cold dark matter model' due to late decay of cold thermal relics into lighter nonrelativistic particles with low phase-space density \cite{Kap05,StrKapBul07}.  Here we consider the relative core radius $x_c=R_c/R$ of DM clumps as a free parameter in the range $0.001-0.1$. Correct approach includes a gradual mass loss of a systems \cite{gnedin2,TayBab01,DieKuhMad}, in particular, by small-scale DM clumps \cite{ZTSH,bde08}.

\section{Formation and Destruction of clumps in hierarchial clustering}
\label{hdestr}

The process of hierarchical clustering and tidal destruction of DM clumps can be outlined in the following way. The DM clumps of minimal mass are formed first in the expanding Universe. The clumps of larger mass, which host the smaller ones are formed later, and so on. Some part of clumps are destroyed tidally in the gravitational field of their host clumps. In this Section we study the destruction of DM clumps in the process of hierarchical structuring long before the final galaxy formation. At small-mass scales the hierarchial clustering is a fast and rather complicated nonlinear process. A formation time of clump with an internal density $\rho$ is $t=(\kappa \rho_{\rm eq}/\rho)^{1/2}t_{\rm eq}$, where $\kappa=18\pi^2$ and $\rho_{\rm eq}=\rho_0(1+z_{\rm eq})^3$ is a cosmological density at the time of matter-radiation equality $t_{\rm eq}$, $1+z_{\rm eq}=
2.35\times10^4\Omega_mh^2$ and $\rho_0=1.9\times10^{-29}\Omega_mh^2\mbox{ g cm}^{-3}$. The index `eq' here and throughout below refers to quantities at the time of matter-radiation equality $t_{\rm eq}$. The DM clumps of mass $M$ can be formed from density fluctuations of different peak-height $\nu=\delta_{\rm eq}/\sigma_{\rm eq}(M)$, where $\sigma_{\rm eq}(M)$ is the fluctuation dispersion on a mass-scale $M$ at the time $t_{\rm eq}$. A mean internal density of clump $\rho$ is fixed at the time of clump formation and according to \cite{cole} equals $\rho= \kappa\rho_{\rm eq}[\nu\sigma_{\rm eq}(M)/\delta_c]^3$, where $\delta_c=3(12\pi)^{2/3}/20\simeq1.686$.

An internal energy of self-gravitating object increases in tidal interactions. This energy increase was calculated, e.\,g., in \cite{gnedin1} for the case of a star globular cluster in a spherical galaxy. By using the model of tidal heating from \cite{gnedin1}, we determine now a survival time (or a time of tidal destruction) $T$ of some chosen small-scale clump due to the tidal heating inside of a larger mass host clump. The motion of a clump would be rather complicated in the case of a fast hierarchical clustering of hosts. During a dynamical time in the host $t_{\rm dyn}\simeq0.5(G\rho_h)^{-1/2}$, where $\rho_h$ is a mean internal density of the host, the chosen small-scale clump can belong to several successively destructed hosts. A clump trajectory in the host experiences successive turns accompanied by the ``tidal shocks'' \cite{spit,gnedin1}. For the considered small-scale clump with a mass $M$ and radius $R$, the corresponding internal energy increase after a single tidal shock is
\begin{equation}
 \Delta E\simeq\frac{4\pi}{3}\,\gamma_1G\rho_hMR^2,
 \label{dele}
\end{equation}
where a numerical factor $\gamma_1\sim1$. Let us denote the number of tidal shocks per dynamical time $t_{\rm dyn}$ by $\gamma_2$. A corresponding rate of clump internal energy growth is $\dot E=\gamma_2\Delta E/t_{\rm dyn}$. A clump is destroyed in the host if its internal energy increase due to tidal shocks exceeds a total energy $|E|\simeq GM^2/2R$. As a result, for a typical time $T=T(\rho,\rho_h)$ of the tidal destruction of a small-scale clump with density $\rho$ inside a more massive host with a density $\rho_h$ we obtain:
\begin{equation}
T^{-1}(\rho,\rho_h)=\dot E/|E|\simeq4\gamma_1\gamma_2
G^{1/2}\rho_h^{3/2}\rho^{-1}.
\end{equation}
It turns out that a resulting mass function of small-scale clumps (see in this Section below) depends rather weakly on the value of $\gamma_1\gamma_2$.

The probability of clump survival, determined as a fraction of the clumps with mass $M$ surviving the tidal destruction in hierarchical clustering, is given by the exponential function $e^{-J}$ with
\begin{equation}
 J\simeq\sum\limits_{h} \frac{\Delta t_h}{T(\rho,\rho_h)}.
 \label{jsum}
\end{equation}
Here $\Delta t_h$ is a difference of formation times $t_h$ for two successive hosts, and summation goes over all clumps of intermediate mass-scales, which successively host the considered small-scale clump of a mass $M$. Changing the summation by integration in (\ref{jsum}) we obtain
\begin{equation}
 J(\rho,\rho_f)=\int\limits_{t_1}^{t_f}\!\frac{dt_h}{T(\rho,\rho_h)}
 \simeq\gamma\frac{\rho_1-\rho_f}{\rho}
 \simeq\gamma\,\frac{\rho_1}{\rho}\simeq\gamma\,\frac{t^2}{t_1^2}\,,
 \label{sumint1}
\end{equation}
where
\begin{equation}
 \gamma=2\gamma_1\gamma_2\kappa^{1/2}G^{1/2}
 \rho_{\rm eq}^{1/2}t_{\rm eq}\simeq14(\gamma_1\gamma_2/3),
 \label{bigj14}
\end{equation}
and $t$, $t_1$, $t_f$, $\rho$, $\rho_1$ and $\rho_f$ are respectively the formation times and internal densities of the considered clump and of its first and final hosts. One may see
from Eq.~(\ref{sumint1}) that the first host provides a major contribution to the tidal destruction of the considered small-scale clump, especially if the first host density $\rho_1$ is close to $\rho$, and consequently $e^{-J}\ll 1$.

Now we need to track the number of clumps $M$ (originated from the density peak $\nu$) which enter some larger host during time intervals $\Delta t_1$ around each $t_1$ beginning from the time $t$ of clump formation. A mass function of small-scale clumps (i.\,e., a differential mass fraction of DM in the form of clumps survived in hierarchical clustering) can be expressed as
\begin{equation}
\label{phiin}
 \xi\frac{dM}{M}\,d\nu=dM\,d\nu\,\frac{e^{-\nu^2/2}}{\sqrt{2\pi}}\!\!
 \int\limits_{t(\nu\sigma_{\rm eq})}^{t_0}\!\!\!dt_1
 \left|\frac{\partial^2 F(M,t_1)}{\partial M~\partial t_1}\right|
 e^{-J(t,t_1)}.
\end{equation}
In this expression $t_0$ is the Universe age and $F(M,t)$ is a mass fraction of unconfined clumps (i.\,e., clumps, not belonging to more massive hosts) with a mass smaller than $M$ at time $t$.
According to \cite{cole}, the mass fraction of unconfined clumps is $F(M,t)={\rm erf}\left( \delta_c/[\sqrt{2}\sigma_{\rm eq}(M)D(t)]\right)$, where ${\rm erf}(x)$ is the error-function
and $D(t)$ is the growth factor normalized by $D(t_{\rm eq})=1$. An upper limit of integration $t_0$ in Eq.~(\ref{phiin}) is not crucial and may be extrapolated to infinity because a main contribution to the tidal destruction of clumps is provided by the early formed hosts at the beginning of the hierarchical clustering. Two processes are responsible for time evolution of the fraction $\partial^2 F/(\partial M\partial t)$ for unconfined clumps in the mass interval $dM$: (i) the formation of new clumps and (ii) the capture of smaller clumps into the larger ones. Both these processes are equally efficient at the time when $\partial^2 F/(\partial M\partial t)=0$.  Finally, we transform the distribution function (\ref{phiin}) to the following form:
\begin{equation}
 \xi\,\frac{dM}{M}\,d\nu\simeq
 \frac{\nu\,d\nu}{\sqrt{2\pi}}\,e^{-\nu^2/2}
 f_1(\gamma)\frac{d\log\sigma_{\rm eq}(M)}{dM}\,dM,
 \label{psiitog}
\end{equation}
where
\begin{equation}
 f_1(\gamma)=
 \frac{2[\Gamma(1/3)-\Gamma(1/3,\gamma)]}{3\sqrt{2\pi}\gamma^{1/3}}.
 \label{f1fun}
\end{equation}
Here $\Gamma(1/3)$ and $\Gamma(1/3,\gamma)$ are the Euler gamma-function and incomplete gamma-function, respectively. The function $f_1(\gamma)$ varies rather slowly in the interesting interval of $14<\gamma<40$, and one may use $f_1(\gamma)\simeq0.2-0.3$. Physically the first factor $\nu$ in (\ref{psiitog}) corresponds to a more effective survival of high-density clumps (i.\,e., with large values of $\nu$) with respect to the low-density ones (with small values of $\nu$). Integrating Eq.~(\ref{psiitog}) over $\nu$, we obtain
\begin{equation}
 \xi_{\rm int}\frac{dM}{M}\simeq0.02(n+3)\,\frac{dM}{M}.
 \label{xitot}
\end{equation}
An effective power-law index $n$ in Eq.~(\ref{xitot}) is determined as $n=-3(1+2\partial\log\sigma_{\rm eq}(M)/\partial\log M)$ and depends very weakly on $M$.

In further calculations we use an interpolation fitting of the fluctuation dispersion $\sigma_{\rm eq}(M)$ from \cite{bugaev} and \cite{SchSchWid99}. The analysis of the WMAP data of the CMB anisotropy \cite{wmap} reveals a  power-law spectrum of initial perturbations with $n_p=0.99 \pm 0.04$ in a good agreement with the canonical inflation value $n_p=1.0$. However, when the data from 2dF galaxy power-spectrum and Ly-$\alpha$ are included in the analysis, the best-fit favors in a softer spectrum with $n_p=0.96 \pm 0.02$.

The simple $M^{-1}$ shape of the mass function (\ref{xitot}) is in a very good agreement with the corresponding numerical simulations \cite{DieMooSta05}, but our normalization factor is a few times smaller. It must be stressed that a physical meaning of the survived clump distribution function $\xi\,d\nu\,dM/M$ is different from the similar one for the unconfined clumps, given by the Press-Schechter mass function $\partial F/\partial M$.

\section{Tidal destruction of clumps by the Galactic disk and stars}
\label{disksec}

Crossing the Galactic disc, a clump can be tidally destructed by the collective gravitational field of stars in the disc. This phenomenon is similar to the destruction of a globular cluster by
the ``tidal shocking'' in the Galactic disc \cite{OstSpiChe}. The kinetic energy gain of a DM particle with respect to the center of clump after one crossing of the Galactic disk is \cite{OstSpiChe}
\begin{equation}
 \delta E=\frac{4g_m^2(\Delta z)^2m}{v_{z,c}^2}A(a),
 \label{egain}
\end{equation}
where $m$ is a constituent DM particle mass, $\Delta z$ is a vertical  distance (orthogonal to the disk plane) of a DM particle with respect to the center of clump, $v_{z,c}$ is a vertical
velocity of clump with respect to the disk plane at the moment of disk crossing and $A(a)$ is the adiabatic correction factor. A gravitational acceleration near the disk plane is $g_m(r)=2\pi G\sigma_s(r)$, where we use an exponential model for a surface density of disk. The factor $A(a)$ in (\ref{egain}) describes the adiabatic protection from slow tidal effects \cite{Wein1}. In \cite{gnedin2} the following fitting formula was proposed: $A(a)=(1+a^2)^{-3/2}$. Here the adiabatic parameter $a=\omega\tau_d$, where $\omega$ is an orbital frequency of DM particle in the clump, $\tau_d\simeq H_d/v_{z,c}$ is an effective duration of gravitational tidal shock produced by the disk with a half-thickness $H_d$.

As a representative example we consider the isothermal internal density profile of DM clump
\begin{equation}
\rho_{\rm int}(r)=\frac{1}{4\pi}\frac{v_{\rm rot}^2}{Gr^2}
 \label{iso}
\end{equation}
with a cutoff at the virial radius $R$: $\rho(r)=0$ at $r>R$. A corresponding mass profile of clump is $M(r)=M_i(r/R)$, where $M_i$ is an initial mass of clump at the epoch of Galaxy formation. With this mass distribution a circular velocity inside a clump is independent of radius, $v_{\rm rot}=(GM(r)/r)^{1/2}= (GM_i/R)^{1/2}$. A gravitational potential corresponding to the density profile (\ref{iso}) is $\phi(r)=v_{\rm rot}^2[\log(r/R)-1]$. Let us define a dimensionless energy of the DM particle $\varepsilon=E/(mv^2_{\rm rot})$ and gravitational potential $\psi(r)=\phi(r)/v^2_{\rm rot}= \ln(r/R)-1$. An internal density profile $\rho_{\rm int}(r)$ and the distribution function of DM particles in the clump $f_{\rm cl}(\varepsilon)$ are related by the integral relation
\cite{Edd16}
\begin{equation}
 \rho_{\rm int}(r)=2^{5/2}\pi\int\limits_{\psi(r)}^{0}
 \sqrt{\varepsilon-\psi(r)}\,f_{\rm cl}(\varepsilon)\,d\varepsilon,
 \label{ferho}
\end{equation}
The corresponding isothermal distribution function is $f_{\rm cl}(\varepsilon)\propto \exp(-2\varepsilon)$.

By using the hypothesis of a tidal stripping of outer layers of a DM clump, we see that a tidal energy gain $\delta\varepsilon$ causes the stripping of particles with energies in the range $-\delta\varepsilon<\varepsilon<0$. A corresponding variation of density at radius $r$ is
\begin{equation}
\delta \rho(r)=2^{5/2}\pi\int\limits_{-\delta\varepsilon}^{0}
 \sqrt{\varepsilon-\psi(r)}\,f_{\rm cl}(\varepsilon)\,d\varepsilon.
 \label{ferho2}
\end{equation}
In this equation the tidal energy gain (\ref{egain}) by different DM particles is averaged over angles, so as $\langle(\Delta z)^2\rangle=r^2/3$. A resulting total mass loss by DM clump during one crossing of the Galactic disk is
\begin{equation}
\delta M=-4\pi\int_0^R r^2\delta\rho(r)\,dr.
\end{equation}
Let us specify the dimensionless quantities
\begin{equation}
 Q_d=\frac{g_m^2}{2\pi v_{z,c}^2G\bar\rho_i}, \quad
 S_d=\frac{4\pi}{3}G\bar\rho_i\tau^2_d,
 \label{eqqq}
\end{equation}
where $\bar\rho_i=3M_i/(4\pi R^3)$ is a initial mean density of clump. For the most parts of clumps $Q_d\ll 1$ with a typical value $Q_d\sim0.03$. The fitting formula for the mass loss of clump during one passage through the Galactic disk is
\begin{equation}
 \left(\frac{\delta M}{M}\right)_d\simeq
 -0.13Q_d\exp\left(-1.58S_d^{1/2}\right).
 \label{mmq}
\end{equation}
Let us choose some particular clump moving in the spherical halo with an orbital ``inclination'' angle $\gamma$ between the normal vectors of the disk plane and orbit plane. The orbit angular velocity at a distance $r$ from the Galactic center is $d\phi/dt=J/(mr^2)$, where $J$ is an orbital angular momentum of a clump. A vertical velocity of a clump crossing the disk is
\begin{equation}
 v_{z,c}=\frac{J}{mr_s}\sin\gamma,
 \label{vzc}
\end{equation}
where $r_s$ is a radial distance of crossing point from the Galaxy center. There are two crossing points (with different values of $r_s$) during an orbital period.

The standard Navarro-Frenk-White profile of the DM Galactic halo is
\begin{equation}
 \rho_{\rm H}(r)=
 \frac{\rho_{0}}{\left(r/L\right)\left(1+r/L\right)^2},
 \label{halonfw}
\end{equation}
where $L=45$~kpc, $\rho_{0}=5\times10^6{\rm M}_\odot$~kpc$^{-3}$. It useful to introduce the dimensionless variables:
\begin{equation}
 x=\frac{r}{L}, \quad \tilde\rho_{\rm H}(x)=\frac{\rho_{\rm
 H}(r)}{\rho_0}, \quad y=\frac{J^2}{8\pi G\rho_0L^4M^2}, \quad
 \varepsilon=\frac{E_{\rm orb}/M-\Phi_0}{4\pi G\rho_0L^2 }, \quad
 \psi=\frac{\Phi-\Phi_0}{4\pi G\rho_0L^2},
\end{equation}
where $\Phi_0=-4\pi G\rho_0L^2$, $E_{\rm orb}$ is a total orbital energy of a clump. With these variables the density profile of the halo (\ref{halonfw}) is written as $\tilde\rho_{\rm H}(x)=[x(1+x)^2]^{-1}$. A gravitational potential $\psi(x)$, corresponding to density profile (\ref{halonfw}) is $\psi(x)=1-\log(1+x)/x$. The relation between density profile $\tilde\rho_H(x)$ and the distribution function is given by the same equation (\ref{ferho}) with an obvious substitution $f_{\rm cl}\Rightarrow F(\varepsilon)$, where the distribution function $F(\varepsilon)$ for a halo profile (\ref{halonfw}) can be fitted as in \cite{Wid00}.

Let us denote the interval of time for motion from $x_{\rm min}$ to $x_{\rm max}$ and back as $T_c(x,\varepsilon,p)$, and the angle of orbital precession during the time $T_c/2$ as $\tilde\phi$. The orbital period is longer than $T_c$ and is given by $T_t=T_c\left(1+ \tilde\phi/\pi\right)^{-1}$. Choosing a time interval $\Delta T$ much longer than a clump orbital period $T_t$, but much shorter than the age of the Galaxy $t_0$, i.\,e.,
$T_t\ll\Delta T\ll t_0$, we define an averaged rate of mass loss by a selected clump under influence of tidal shocks in successive disk crossings:
\begin{equation}
\frac{1}{M}\left(\frac{dM}{dt}\right)_d\simeq\frac{1}{\Delta
T}\sum \left(\frac{\delta M}{M}\right)_d,
 \label{deriv2}
\end{equation}
where $(\delta M/M)_d$ is given by (\ref{mmq}) and summation goes over all successive crossing points (odd and even) of a clump orbit with the Galactic disk during the time interval $\Delta T$. The values of the $g_m$ and $v_{z,c}$ both depend on the radius $x=r/L$. One simplification in calculation of (\ref{deriv2}) follows from the fact that a velocity of orbit precession is constant. For this reason the points of successive odd crossings are separated by the same angles $\tilde\phi$. The same is also true for successive even crossings. Using this simplification we transform the summation in (\ref{deriv2}) to integration:
\begin{eqnarray}
 \frac{1}{\Delta T} \sum\left(\frac{\delta M}{M}\right)_d
 \simeq\frac{2}{T_t|\tilde\phi|}\int\limits_{x_{\rm min}}^{x_{\rm max}}
 \left(\frac{\delta M}{M}\right)_d\frac{d\phi}{dx}dx,
 \nonumber
\end{eqnarray}
where $d\phi/dx$ is an equation for clump orbit in the halo.

During a single close encounter of a DM clump with a star, the energy gain of a constituent DM particle in the clump with respect to clump center is \cite{bde06}:
\begin{equation}
 \delta E=\frac{2G^2m_s^2m\Delta z^2}{v_{\rm rel}^2l^4},
 \label{egainstar}
\end{equation}
where $m_*$ is a star mass, $l$ is an impact parameter, $v_{\rm rel}$ is a relative star velocity with respect to a clump, $\Delta z=r\cos\psi$, $r$ is a radial distance of a DM particle from the clump center and $\psi$ is an angle between the directions from the clump center to the DM particle and to the point of closest approach of a star. Using the same method as in the Section~\ref{disksec} we calculate a relative mass loss by clump $(\delta M/M)_s$ during a single encounter with a star and obtain the same fitting formula as (\ref{mmq}) but with a substituting the dimensionless parameters, $Q_d\Rightarrow Q_s$ and $S_d\Rightarrow S_s$, where
\begin{equation}
 Q_s=\frac{Gm_*^2}{2\pi v_{\rm rel}^2l^4\bar\rho_i},
 \quad
 S_s=\frac{4\pi}{3}G\bar\rho_i\tau^2_s,
 \label{eqqq2}
\end{equation}
where $\tau_s\simeq l/v_{\rm rel}$.

A DM clump acquires the maximum energy gain during a single encounter with a star when impact parameter $l\sim R$. Integrating over all impact parameters $l>R$, we calculate an averaged rate of mass loss by a clump during successive encounters with stars:
\begin{equation}
 \label{mmqstar}
\frac{1}{M}\left(\frac{dM}{dt}\right)_s\simeq
 \frac{1}{2T_t\sqrt{2\pi G\rho_0}} \int\limits_{R}^{\infty}\!2\pi
 l\,dl\!\int\limits_{x_{\rm min}}^{x_{\rm max}}\!\!
 \frac{ds\, n_*(s)v_{rel}}{\sqrt{\varepsilon-\psi(s)-y/s^2}}
 \left(\frac{\delta M}{M}\right)_{\!\!s}\!,
\end{equation}
where $n_*(r)$ is a radial number density distribution of stars in the bulge and halo.

\section{Surviving fraction of clumps}
\label{secsurv}

From Eq.~(\ref{egain}) it is seen that the tidal forces influence mainly the outer part of clump (where $\Delta z$ is rather large). Further we use our basic assumption that only outer layers of a clump undergo the tidal stripping, while the inner parts of a clump are unaffected by tidal forces. Thus we assume that a clump mass $M=M(t)$ and radius $R=R(t)$ are both diminishing in time due to the tidal stripping of outer layers but its internal density profile remains the same as given by Eq.~(\ref{iso}), e.\,g., for the isothermal density profile $M(t)\propto R(t)$ and $\bar\rho(t)\propto M(t)^{-2}$. Combining together the rates of mass loss (\ref{deriv2}) and (\ref{mmqstar}) due to the tidal stripping of a clump by the disk and stars respectively, we obtain the evolution equation for a clump mass:
\begin{equation}
 \frac{dM}{dt}=
 \left(\frac{dM}{dt}\right)_d+\left(\frac{dM}{dt}\right)_s
 \label{mmtmain}.
\end{equation}
In the following we solve this equation numerically starting from the time of Galaxy formation at $t_0-t_G$ up to  the present moment $t_0$.

The most important astrophysical manifestation of DM clumps is a possible annihilation of the constituent SUSY DM particles. The crucial point is a dominance of the central core of a clump in annihilation signal if clumps have a steep enough density profile. Namely, annihilation of DM particles in a clump core prevail in a total annihilation rate in a single clump with a power-law density profile (\ref{rho}) if $\beta>3/2$ and $x_c=R_c/R\ll1$. More specifically, the quantity $\dot N\propto\int_{r_0}^{r}4\pi r'^2dr'\rho_{\rm int}^2(r')$ practically does not depend on $r$, if $r \gg r_0$. As a result the annihilation luminosity of DM clump with approximately isothermal density profile ($\beta\simeq 2$) will be nearly constant under influence of tidal stripping until a clump radius diminishes to its core radius. In other words, in the nowadays Galaxy the remnants of tidally stripped clumps with $x_c<\mu(t_0)\ll1$, where $\mu(t)=M(t)/M_i$ and $t_0\simeq10^{10}$~yrs is the Galaxy age, obeys the evolution equation (\ref{mmtmain}) and have the same annihilation luminosity as their progenitors with $\mu=1$.

By using evolution equation (\ref{mmtmain}) we now calculate the probability $P$ of the survival of clump remnant during the lifetime of the Galaxy. Let us choose some arbitrary point in the halo with a radius-vector $\vec r$ and an angle $\alpha$ with a polar axis of the Galactic disk. Only the clump orbits with inclination angle $\pi/2-\alpha<\gamma<\pi/2$ pass through this point. A survival probability for clumps can be written now in the following form
\begin{equation}
 \label{sp1}
 P(x,\alpha)=\frac{4\pi\sqrt{2}}{\tilde\rho(x)\sin\alpha}
 \int\limits_{0}^{1}dp\int\limits_{0}^{\sin\alpha}d\cos\gamma
 \!\int\limits_{\psi(x)}^{1}\!d\varepsilon\,
 [\varepsilon-\psi(x)]^{1/2}F(\varepsilon){\rm \Theta}[\mu(t_0)-x_c].
  \end{equation}
In this equation $\tilde\rho(x)$ is a density profile of the halo from (\ref{halonfw}), $p=\cos\theta$, $\theta$ is an angle between the radius-vector $\vec r$ and the orbital velocity of clump, ${\rm \Theta}$ is the Heaviside function, $\psi(x)$ is the halo gravitational potential from (\ref{halonfw}), $F(\varepsilon)$ is a distribution function of clumps in the halo, $\mu(t_0)$ depends on all variables of the integration and $x_c=R_c/R$ is an initial value of clump core. The function $\mu(t_0)$ is calculated from numerical solution of evolution equation (\ref{mmtmain}). If $\mu(t_0)>x_c$, the clump remnant is survived through the tidal destruction by both the disk and stars. The annihilation rate in this remnant would be the same as in the initial clump. On the contrary in the opposite case, when $\mu(t_G)<x_c$, the clump is totally destructed because (i) the core is not a dynamically separated system and composed of particles with extended orbits, and because (ii) a nearly homogeneous core is destructed easier than a similar object with the same mass but with a near isothermal density profile.

According to theoretical model \cite{bde03} and  numerical simulations \cite{DieMooSta05}, a differential number density of small-scale clumps  in the co-moving frame in the Universe is $n(M)\,dM \propto dM/M^2$. This distribution is calculated in \cite{bde06}. The damping of small-scale perturbations with $M<M_{\rm min}$ provides an additional factor $\exp[-(M/M_{\min})^{2/3}]$ responsible for the fading of distribution at small $M$. The result of numerical simulations \cite{DieMooSta05} can be expressed in the form of a differential mass fraction of the DM clumps in the Galactic halo $f(M)\,dM\simeq \kappa(dM/M)$, where $\kappa\simeq8.3\times10^{-3}$. The analytical estimation (\ref{xitot}) gives approximately $\kappa\simeq4\times10^{-3}$ for the mass interval $10^{-6}{\rm M}_\odot<M<1{\rm M}_\odot$. The discrepancy by the factor $\simeq2$ may be attributed to the approximate nature of our approach as well as to the well known additional factor 2 in the original Press--Schechter derivation of the mass function. In the latter case one must simply multiply equation (\ref{phiin}) by factor 2.

\section{Amplification of annihilation signal}

A local annihilation rate is proportional to the square of DM particle number density. A number density of DM particles in clump is much large than a corresponding number density of the diffuse (not clumped) component of DM. For this reason an annihilation signal from even a small fraction of DM clumps can dominate over an annihilation signal from the diffuse component of DM in the halo. We consider here the Harrison-Zeldovich initial perturbation spectrum with power index $n_p=1$ as a representative  example. The value of $n_p$ is not exactly fixed by the current observations of CMB anisotropy. In the case of $n_p<1$, the DM clumps are less dense, and a corresponding amplification of annihilation signal would be rather small \cite{bde03}.

The gamma-ray flux from annihilation of diffuse distribution (\ref{halonfw}) of DM in the halo is proportional to
\begin{equation}
 I_{\rm H}=\int\limits_{0}^{r_{\rm max}(\zeta)}\rho_{H}^2(\xi)\,dx,
 \label{ihal1}
\end{equation}
where the integration is over $r$ goes along the line of sight, $\xi(\zeta,r) = (r^2+r_{\odot}^2-2rr_{\odot}\cos\zeta)^{1/2}$ is the distance to the Galactic center, $r_{\rm max}(\zeta) = (R_{\rm H}^2-r_{\odot}^2\sin^2\zeta)^{1/2} + r_{\odot}\cos\zeta$  is a
distance to the external halo border, $\zeta$ is an angle between the line of observation and the direction to the Galactic center, $R_{\rm H}$ is a virial radius of the Galactic halo,
$r_{\odot}=8.5$~kpc is the distance between the Sun and Galactic center. The corresponding signal from annihilations of DM in clumps is proportional to the quantity \cite{bde03}
\begin{equation}
 I_{\rm cl}= S\int\limits_{0}^{r_{\rm max}(\zeta)}\!dx\!\!
 \int\limits_{M_{\rm min}}\!\!f(M)\,dM
 \rho\rho_{H}(\xi)P(\xi,\rho),
 \label{ihal2}
\end{equation}
where $\rho(M)$ is the mean density of clump. The function $S$ depends on the clump density profile and core radius of clump \cite{bde03} and we use $S\simeq 14.5$ as a representative example. The observed amplification of the annihilation signal is defined as $\eta(\zeta)= (I_{\rm cl}+I_{\rm H})/I_{\rm H}$ is calculated numerically in \cite{bde03,bde06} for the case $x_c=0.1$. It tends to unity at $\zeta\to 0$ because of the divergent form of the halo profile (\ref{halonfw}). The annihilation of diffuse DM prevails over signal from clumps at the the Galactic center. The observed signal is obtained by integration along the line of sight and the effect of clumps destruction at the Galactic center is masked by the signal from another regions of the halo. This amplification of an annihilation signal is called a ``boost-factor'' \cite{deBoer}.

\section{Anisotropy of the observed signal}

The usual assumption in calculations of DM annihilation is a spherical symmetry of the Galactic halo. In this case an anisotropy of annihilation gamma-radiation is only due to off-center position of the Sun in the Galaxy. Nevertheless, a principal significance of the halo nonsphericity for the observed annihilation signal was demonstrated in \cite{CalMoo00}. According to observations, the axes of the Galactic halo ellipsoid differ most probably no more than $10-20$\%, but even a much more larger difference of axes, up to a factor 2, can not be excluded \cite{OllMer00,OllMer01}. This leads to more than an order of magnitude uncertainty in the predicted annihilation flux from the Galactic anti-center direction \cite{CalMoo00}. It was shown \cite{bde03,bde06} that (i) small-scale DM clumps dominate in the generation of annihilation signal and (ii) the Galactic stellar disk provides the main contribution to the tidal destruction of clumps at $r>3$~kpc, i.\,e., outside the central bulge region. The detectors at the Fermi satellite are sensitive to anisotropy up to $0.1$\% level \cite{HooSer07}. This provide a hope to discriminate the anisotropic DM annihilation signal from the diffuse gamma-ray backgrounds. To calculate the anisotropy we used a simplified criterium for a tidal destruction of clump: clump is destructed if a total tidal energy gain $\sum(\Delta E)_j$ after several tidal shocking becomes of order of initial binding energy of a clump $|E|$, i.\,e., $\Delta E=\sum\limits_j(\Delta E)_j\sim|E|$, where summation goes over the successive disk crossings (or encounters with stars). In this model the survival probability for clumps can be written now in the following form
\begin{equation}
 P(r,\alpha)\!=\!\frac{4\pi\sqrt{2}}{\tilde\rho(x)\sin\alpha}
 \int\limits_{0}^{1}dp\int\limits_{0}^{\sin\alpha}d\cos\gamma
 \int\limits_{\psi(x)}^{1}\!d\varepsilon\,
 [\varepsilon-\psi(x)]^{1/2}F(\varepsilon)\,e^{-\Delta E/|E|},
 \label{sp2}
\end{equation}
where the definitions are the same as in (\ref{sp1}) except for exponential factor for clumps destruction $e^{-\Delta E/|E|}$. In \cite{bde03} the triple integral from (\ref{sp2}) for survival probability $P(r,\alpha)$ was numerically calculated. As a representative example we consider the Earth-mass clumps $M=10^{-6}M_{\odot}$ originated from $2\sigma$ density peaks in the case of power-law index of primordial spectrum of perturbations $n_p=1$. The mean internal density of these clumps is $\rho\simeq7\times10^{-23}$~g~cm$^{-3}$. The annihilation signal for the Galactic disk plane and for the orthogonal vertical plane (passing through the Galactic center) as a function of angle $\zeta$ between the observation direction and the direction to the Galactic center was calculated numerically in \cite{bde07}. The difference of the signals in two orthogonal planes at the same $\zeta$ can be considered as an anisotropy  measure. Defined as $\delta=(I_2-I_1)/I_1$, it has a maximum value $\delta \simeq 0.09$ at  $\zeta\simeq39^{\circ}$.

\section{Conclusion}

We demonstrate that the cores of the small-scale DM clumps (or clump remnants) survive  in general during clump formation and their tidal destruction by stars in the Galaxy. These small-scale DM clumps may provide the major contribution to the annihilation signal (in comparison with the diffuse DM) in the Galactic halo. The amplification (boost-factor) can reach $10^2$ or even $10^3$ depending on the initial perturbation spectrum and minimum mass of clumps.

\end{document}